\documentclass[letter]{aa}
\usepackage{graphicx}
\usepackage{txfonts}
\begin{document}
   \title{Intra Cluster Globular Clusters around NGC\,1399 in Fornax?}
   \author{Y. Schuberth
     \inst{1,2}
        \and
     T. Richtler
     \inst{1}
     \and
    L. Bassino     
     \inst{3}
     \and
     M. Hilker
     \inst{4}
}

   \offprints{Y.~Schuberth (\texttt{ylva@astro.uni-bonn.de})}
\institute{ Universidad de Concepci\'on, Departamento de F\'isica, Casilla
     160-C, Concepci\'on, Chile 
       \and
Argelander-Institut f\"ur Astronomie,
     Universit\"at Bonn, Auf dem H\"ugel 71, D-53121 Bonn, Germany   
      \and 
Facultad de Ciencias Astron\'omicas y Geof\'isicas, Universidad Nacional de La Plata, Paseo del Bosque S/N, 1900--La Plata, Argentina and IALP-CONICET
  \and European Southern Observatory,
     Karl-Schwarzschild-Str.~2, D-85748 Garching, Germany 
}

   \date{Received 14 September, 2007; accepted 16 October, 2007}

  \abstract
{}
  { We investigate whether the globular clusters (GCs) in the recently published sample of  
GCs in the Fornax cluster by Bergond and coworkers are indeed intra--cluster objects.}
   {We combine the catalogue of radial velocity measurements by Bergond et al.~with our CTIO MOSAIC 
photometry in the  Washington system  and analyse the relation of metal--poor and metal--rich GCs with their  host galaxies.  }
{The metal--rich GCs appear to be kinematically  associated with their respective host galaxies. The vast majority of the metal--poor GCs found in between the galaxies of the Fornax cluster have velocities which are  consistent with them being members of the very extended NGC\,1399 GC system. { We find  that when the sample is restricted to the most accurate velocity measurements, the GC velocity dispersion profile can be described with a mass model derived for the NGC\,1399 GC system within 80\,kpc. We identify one ``vagrant'' GC whose radial velocity suggests that it is not bound to any galaxy unless its orbit has a very large apogalactic distance.}}
      {}
{}

   \maketitle
%

\section{Introduction}
The existence of intra--cluster globular clusters (ICGCs), i.e.
globular clusters (GCs) which are not bound to individual galaxies
but, rather, move freely through the potential well of a galaxy
cluster as a whole, was proposed by  White~(\cite{white87}) and
West et al.~(\cite{west95}). In the nearby
($D=19\,\rm{Mpc}$) Fornax cluster of galaxies, ICGC candidates were
identified as an excess population of GC candidates in the vicinity of
dwarf galaxies (Bassino et al.~\cite{bassino03}), and as a region of
enhanced number density of GC candidates in between the central
dominant galaxy NGC\,1399 and its second--nearest (giant) neighbour,
NGC\,1387 (Bassino et al.~\cite{b06a}, cf.~their Fig.~9).  Tamura et
al.~(\cite{tamura06}) performed a wide--field survey of the central
Virgo cluster and detected an excess population of GC candidates far
away from any major galaxy. The spectroscopic confirmation of these
candidate ICGCs, however, is still pending.
\par Recently, Bergond et al.~(\cite{bergond07}) presented the
velocities of a sample of GCs in the Fornax cluster, with some objects
having projected distances of more than 230\,kpc from NGC\,1399. These
authors  labelled a significant fraction of their objects as ICGCs.
Yet, one has to bear in mind that, with the photometric study by
Bassino et al.~(\cite{b06a}), it became clear that the very populous
globular cluster system (GCS) of the central galaxy, NGC\,1399, has an
extent of at least 250\,kpc, which is comparable to the core radius of
the cluster ($R_{King}=0.7^{\circ}\simeq 230\,\rm{kpc}$, Ferguson
\cite{ferguson89}). \par In this \emph{Letter}, we present Washington
photometry for 116 of the 149 GCs of the Bergond et al.~kinematic
sample and demonstrate that the photometric and dynamical properties
of the ``ICGCs'' are rather consistent with those of GCs belonging to
NGC\,1399.
\section{The Data Set}
Below, we give a brief description of the two data sets we combined
for this study. The  table is available in electronic form.
\subsection{The Kinematic Data}
Bergond et al.~(\cite{bergond07}; hereafter B07+) used the FLAMES
multi--object fibre--fed spectrograph in the GIRAFFE/MEDUSA mode on
the VLT to obtain medium--resolution spectra of 149 GCs in the central
region of the Fornax cluster (see their Fig.~1 for the location of the
objects). They determined the velocities via Fourier
cross--correlation, and assigned each measurement a quality flag.
``Class\,A'' indicates a very secure measurement, while ``Class\,B''
measurements have larger uncertainties (see B07+ for details). These
authors define as ICGCs those objects more than $1.5\,d_{25}$ away
from any bright Fornax galaxy\footnote{$d_{25}$ is the isophotal
 diameter of a galaxy at the level of $25\,\rm{mag}/\rm{arcsec}^{2}$ in
the $B$-band.}. In the following, we will use the term ICGC in this
\emph{geometrical} sense, unless otherwise stated.
\subsection{The Photometric Data}
\begin{table}
\caption{The combined data set: GC colours and environment. The
numbers for the full sample, the ICGCs, and the GCs in the masked
areas (see text for details) are given below. The first column gives
the number of GCs with velocity measurements from the B07+
catalogue. The number of red and blue GCs is given in columns 2 and 3,
respectively. Column 4 lists the ratio of blue to red GCs. In all
columns, the number of ``Class\,A'' velocity measurements is given in
parentheses.}  \centering
\begin{tabular}{lrrrrrrrr}\hline \hline
& \multicolumn{2}{c}{B07+}
& \multicolumn{2}{c}{Red}
& \multicolumn{2}{c}{Blue}
& \multicolumn{2}{c}{$N_{\rm{Blue}}/N_{\rm{Red}}$}\\ 
& \multicolumn{2}{c}{(1)}
& \multicolumn{2}{c}{(2)}
& \multicolumn{2}{c}{(3)}
& \multicolumn{2}{c}{(4)}
\\\hline
All & 149 & (109) & 43 & (35) & 73 & (56) &1.7&(1.6)\rule[-1.2ex]{0ex}{2ex} \\ 
ICGC & 61 & (45) & 11 & (9) & 45 & (36)&4.1&(4.0) \\
$R_{\rm{pro}} \leq 1.5d_{25}$ & 88 & (64) & 32 & (26) & 28 & (20)&0.9&(0.8) \\ 
\hline \hline
\end{tabular}
\label{tab:data}
\end{table}
As part of our programme to study the GCSs of
ellipticals in the Fornax cluster, wide--field photometry in the
metallicity--sensitive Washington system was obtained for several
fields, using images from the CTIO MOSAIC camera which has a
field--of--view of $36\arcmin\times36\arcmin$.  The results have been
presented in a series of papers (Dirsch et al.~\cite{dirsch03},
Bassino et al.~\cite{b06a}, and Bassino et al.~\cite{b06b}) to which we
refer the reader for details of the observations and data reduction.
We  use these data to determine the magnitudes and $C\!-\!R$
colours for the GCs presented by B07+.
\subsection{The Combined Data Set}
In total, 121 out of the 149 B07+ GCs (listed in their Table\,1) were
matched to objects in our photometry database (the remaining objects
mostly lie in chip gaps of the undithered MOSAIC images or are too
close to the centres of galaxies). Of these 121 objects, all but five
have colours that lie in the range  $0.8\,\leq C\!-\!R \leq2.3$,
which was used for the GC candidate selection by Dirsch et
al.~(\cite{dirsch03}).  Discarding the objects with deviant colours,
we thus define a sample of 116 GCs which have reliable photometry. Out
of these, 56 are ICGCs according to the definition of B07+. The
remaining 60 GCs are ``masked objects'', i.e.~they are found within
$1.5\,d_{25}$ of a Fornax galaxy (see Fig.~1 of B07+).  \par
\begin{figure}
\centering
\includegraphics[width=0.46\textwidth]{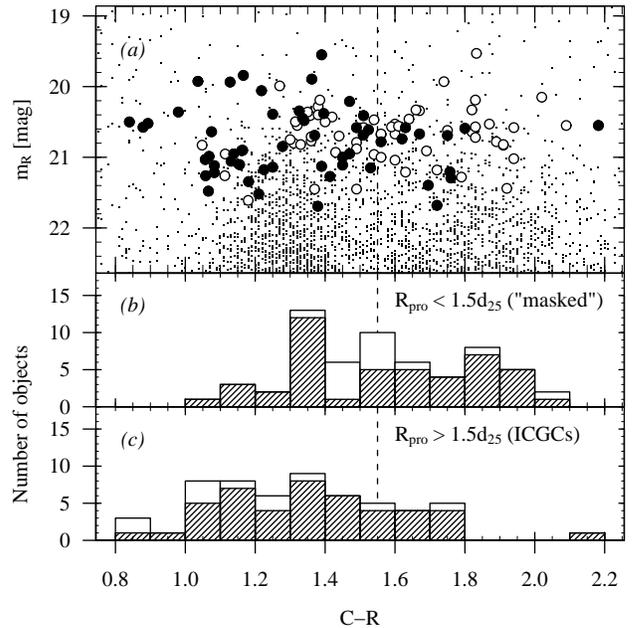}
\caption{MOSAIC photometry. The \emph{upper panel} shows the colour
magnitude diagram for the B07+ objects with MOSAIC photometry. Filled
circles are ICGCs, open circles represent GCs within $1.5\,d_{25}$ of
a giant Fornax member. The objects from the photometric
catalogue of NGC\,1399 GC candidates (central field) by Dirsch et
al.~(\cite{dirsch03}) are shown as dots. The \emph{middle} and
\emph{bottom panel} show the colour histograms for the ``masked''
objects and the ICGCs, respectively. In both panels, the dashed
histograms show the objects with ``Class\,A'' velocity
measurements. The dashed line at $C\!-\!R=1.55$ shows the colour
adopted to separate blue from red GCs (cf.~Dirsch et
al.~\cite{dirsch03}).}
\label{fig:cmd}
\end{figure}

 The upper
panel of Fig.~\ref{fig:cmd} shows the colour--magnitude diagram for
all B07+ objects with reliable Washington photometry (large
symbols). Filled circles indicate GCs classified as ICGCs, and the
dots show the photometry of the NGC\,1399 GCS presented by Dirsch et
al.~(\cite{dirsch03}). Following these authors, we adopt $C\!-\!R=1.55$
(dashed line) as the colour dividing red (metal--rich) from blue
(metal--poor) GCs.  Note that four ICGCs have extremely blue colours
($C\!-\!R \leq 1.0$). The middle and bottom panels of
Fig.~\ref{fig:cmd} show the colour histograms for the ``masked
objects'' and the ICGCs, respectively. One sees that the ICGCs are
preferentially blue, while the GCs in the vicinity of the Fornax
galaxies have a broad distribution of colours.  Table\,\ref{tab:data}
lists the number of red and blue GCs with reliable photometry found in
the two different environments.
\section{Colours and Kinematics of the Globular Clusters}
\begin{figure}
\centering
\includegraphics[width=0.46\textwidth]{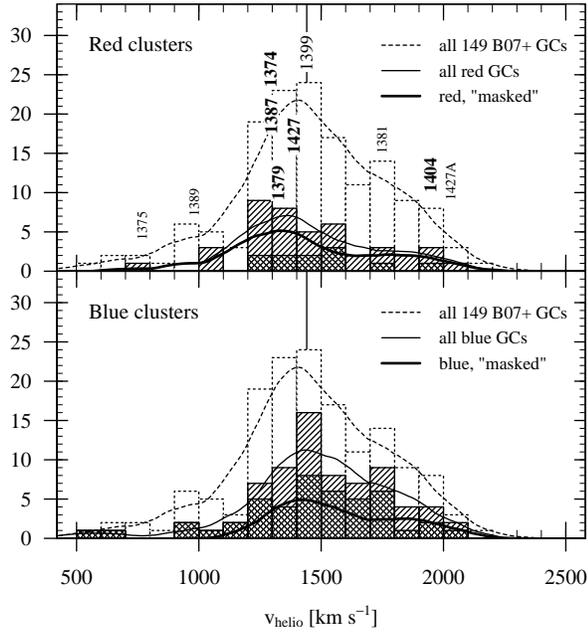}
\caption{\emph{Upper panel:} Velocities for all red GCs (dashed
 histogram) and red ICGCs (hashed). \emph{Lower panel:} The same for
 the blue GCs. For reference, the distribution of all 149 GC
 velocities from the B07+ catalogue is shown as dotted histogram. The
 solid line is the systemic velocity of NGC\,1399.  The curves show
 Epanechnikov kernel density estimates for the samples indicated in
 the legend. In the upper panel, the systemic velocities of the 10
 Fornax galaxies in the area covered by the B07+ study are shown by
 their corresponding NGC numbers.  Boldface labels indicate the five
 brightest galaxies besides NGC\,1399. }
\label{fig:velhist}
\end{figure}
{Figure\,\ref{fig:velhist} compares the velocity distributions of
the blue and red GCs. For comparison, the histogram of all 149 B07+
velocities is shown. The blue GCs (lower panel) show a clear peak near
the velocity of NGC\,1399. The velocity distribution of the red GCs
(upper panel) peaks at a lower velocity.  Note that of the five
brightest galaxies in the survey area, besides NGC\,1399, four have
velocities that are lower than the NGC\,1399 systemic velocity, while
only one (NGC\,1404) has a higher velocity.  } Below, we study the
kinematics of the GCs in the masked areas. The kinematics of the ICGCs
are discussed in Sect.~\ref{subsect:1399icgcs}.
\subsection{The Globular Clusters in the Vicinity of Fornax Galaxies}
\begin{figure}
\centering
\includegraphics[width=0.44\textwidth]{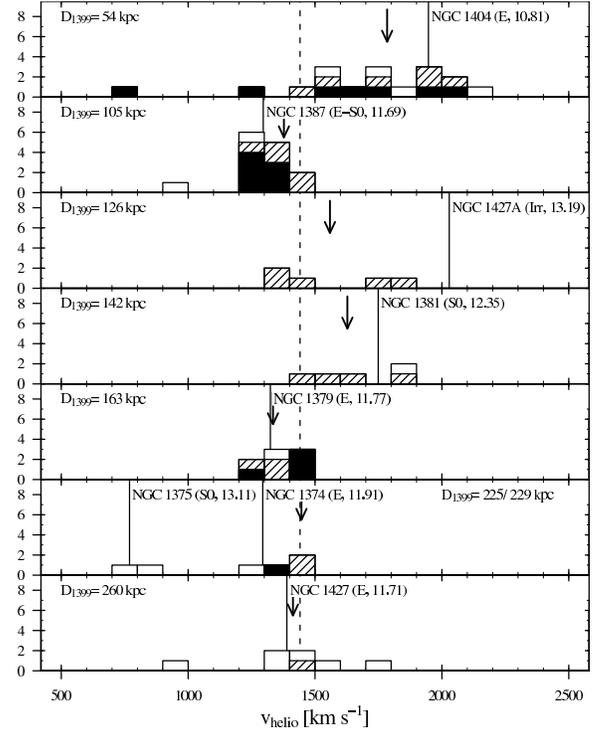}
\caption{Kinematics of the ``masked'' GCs.  In each panel, the
systemic velocity (solid lines) of the Fornax member is shown together
with the velocities of the objects within the corresponding
$3\,d_{25}$ diameter mask as applied by B07+. Solid and hashed bars
represent red and blue GCs, respectively. Hollow histograms show the
objects for which no MOSAIC photometry is available. Arrows indicate
the mean velocity of the blue clusters in the panel. The galaxy type
and the total apparent corrected $B$--magnitude as listed in the
HyperLeda (Paturel et al.~\cite{paturel03}) database are given in
parentheses.  In all panels, the systemic velocity of NGC\,1399 is
shown as dashed line, and $D_{1399}$ denotes the projected distance
from NGC\,1399.  For NGC\,1389 no GCs were observed within its mask
diameter.}
\label{fig:bergondbelong}
\end{figure}

In total, 88 GCs are found in the masked areas centred on NGC\,1399
and nine other Fornax galaxies. In Fig.~\ref{fig:bergondbelong} we
show the velocity distributions of the latter (for NGC\,1399, refer to
Sect.~\ref{subsect:1399icgcs}). Each plot displays the velocities for
the objects covered by the $3\,d_{25}$--diameter mask of the
corresponding Fornax member. The panels are sorted (from top to
bottom) by projected distance from NGC\,1399. In all panels, the
vertical dashed line indicates the systemic velocity of NGC\,1399, and
the velocity of the galaxy is shown as solid line. The arrows indicate
the mean velocity of the blue GCs in a given panel. \par We observe
that, with the exception of NGC\,1404 whose GCS is projected onto that
of NGC\,1399 (and maybe even interacts with NGC\,1399, e.g.~Bekki et
al.~\cite{bekki03} and references therein), the red GCs, if present,
are always found within $100\,\rm{km\,s}^{-1}$ of the host
galaxy. Further we note that all red GCs are associated with bright
($L_B\leq-19.5$) early--type galaxies\footnote{This is probably a
consequence of the limiting magnitude of $V\simeq22.2$. The
spectroscopic survey only probes the brightest part of the GC
luminosity function, and the fainter galaxies, hosting less populous
GCSs, are simply less likely to possess a large number of bright
GCs.}.  \par The velocities of the blue GCs, on the other hand, are
not so strongly correlated with the systemic velocities of the
galaxies. In fact, the velocity distributions of the blue GCs appear
to be shifted towards the systemic velocity of NGC\,1399, as can be
seen from the arrows displayed in the panels. This effect is strongest
for NGC\,1381 and NGC\,1427A which both have high relative velocities
with respect to NGC\,1399. We suggest that the majority of the blue
GCs observed in the vicinity of these two galaxies are  associated
with NGC\,1399. For the masks covering NGC\,1387, NGC\,1379 and
NGC\,1427 it is not possible to  make such a distinction
since the velocities of these galaxies are within just
$150\,\rm{km\,s}^{-1}$ of the NGC\,1399 systemic velocity.\par
\subsection{NGC\,1399 and the ICGCs}
\label{subsect:1399icgcs}
The bottom panel of Fig.~\ref{fig:sigma} shows the velocities of all
objects labelled as ICGCs by B07+ together with the GCs within the
$3\,d_{25}$ area of NGC\,1399 but excluding those within the mask
centred on NGC\,1404. The horizontal and vertical dashed lines
indicate the systemic velocity ($1441\,\rm{km\,s}^{-1}$) and the
$1.5\,d_{25}$ radius of NGC\,1399, respectively.\par The majority of
the ICGCs is blue,  only $20\%$ of the ICGCs are red
(cf.~Table\,\ref{tab:data}). This is in accordance with the steeper
slope of the number density profile found for the red GCs (Bassino et
al.~\cite{b06a}). One further notes that the red ICGCs (``Class\,A'')
have velocities within less than $280\,\rm{km\,s}^{-1}$ of the
systemic velocity of NGC\,1399. Their mean velocity is
$1442\pm52\,\rm{km\,s}^{-1}$, and the dispersion is apparently very
low (for the 9 red ``Class A'' ICGCs, we find
$\sigma_{\rm{los}}=155\pm36\,\rm{km\,s}^{-1}$, but this value should
be treated with caution, given the low number of data points). The
presence of red GCs at large distances from NGC\,1399 is not
unexpected since Bassino et al.~found the red subpopulation of
NGC\,1399 GCs to extend to about 35\arcmin{}. The kinematics of the
red ICGCs suggest that these objects are indeed part of the NGC\,1399
GCS.\par The blue ICGCs within 30\arcmin{} ($\simeq 165\,\rm{kpc}$) of
NGC\,1399 are also clearly concentrated towards the systemic velocity
of the central galaxy, albeit with a larger scatter, indicative of a
higher velocity dispersion.  For the 25 blue ``Class\,A'' ICGCs in this
radial range, we find a mean velocity of $1474\pm53\,\rm{km\,s}^{-1}$,
and a dispersion of $\sigma_{\rm{los}}=266\pm37\,\rm{km\,s}^{-1}$.
\section{Dynamics of the ICGCs}
Bergond et al.~note that the velocity dispersion profile of the ICGCs
exhibits a sharp rise beyond $\sim\!150\,\rm{kpc}$, almost reaching
the value found for the dwarf galaxies in Fornax
($\sigma_{\rm{dwarfs}}=429\pm 41\,\rm{km\,s}^{-1}$, Drinkwater et
al.~\cite{drinkwater01}). This finding suggests, in their view, a
dynamic link between GCs and dwarf galaxies, supporting the
interpretation as intra--cluster objects.  It is interesting to review
this statement in the context of the dynamics of the NGC\,1399 GCS out
to 80\,kpc (Richtler et al.~\cite{r04}, Schuberth et al.~in prep).
\par
\begin{figure}
\centering
\includegraphics[width=0.48\textwidth]{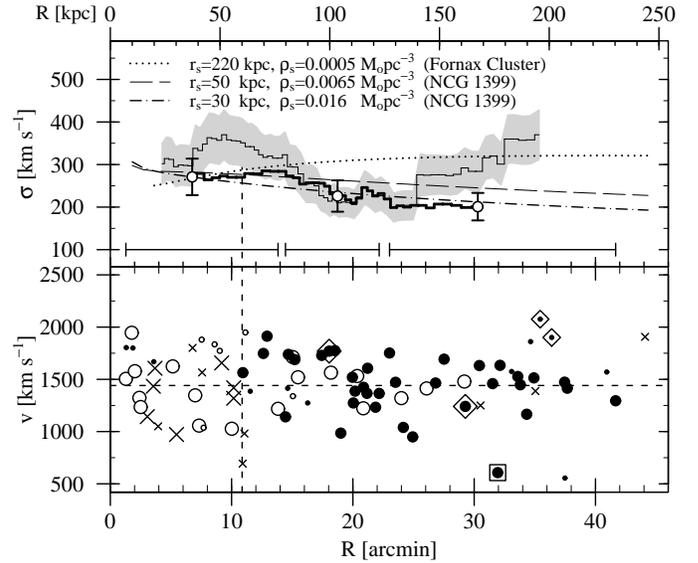}
\caption{\emph{Upper panel:} Line--of--sight velocity dispersion
profiles. The  thin solid line is the dispersion profile for all
objects displayed in the lower panel,  and the grey region shows
the uncertainties. The thick solid line is the profile obtained when
omitting all ``Class\,B'' objects and the most extreme ``Class\,A''
ICGC (indicated by a square in the lower panel).  Both profiles were
calculated using a moving bin containing 20 objects.  The data points
show the results for three independent bins, containing 20 GCs
each. The range of radial distances of the GCs used in a given bin are
indicated by the horizontal bars.  The dashed and dot--dashed
curves are two models derived from the NGC\,1399 GC dynamics within
80\,kpc (see text for details). The dotted curve corresponds to a halo
which reproduces the mass profile of the Fornax cluster as derived by
Drinkwater et al.~(\cite{drinkwater01}), (their Fig.~4, solid
line). \emph{Lower panel:} Velocity vs.~projected distance from
NGC\,1399  for 87 GCs ( GCs within $1.5\,d_{25}$ of other
Fornax galaxies are not plotted). Large and small symbols are
objects with ``Class\,A'' and ``Class\,B'' velocity measurements,
B07+, respectively. Red and blue GCs are shown as open and filled
circles, respectively. Crosses are objects for which no MOSAIC
photometry is available. The four objects with unusually blue colours
($C\!-\!R \leq 1.0$) are marked with diamonds. The square marks the
``Class\,A'' ICGC with the most extreme velocity. In both panels, the
vertical dashed line shows the radius of the $3\,d_{25}$--diameter
mask ($1.5\,d_{25}=10\farcm9 \simeq 60\,\rm{kpc}$) of NGC\,1399.  }
\label{fig:sigma}
\end{figure}

 The  dashed and dash--dotted curves in the upper panel of
Fig.~\ref{fig:sigma} are two isotropic Jeans models derived from the
kinematics of a sample of GCs within 80\,kpc (Richtler et
al.~\cite{r07}, Schuberth et al.~in prep), while the dotted curve
corresponds to a halo which provides a good approximation to the mass
profile of the Fornax cluster as given by  Drinkwater et
al.~(\cite{drinkwater01}). All models were calculated for the sum of
the stellar mass of NGC\,1399 and an NFW halo, {i.e.}  {
$\varrho_{\rm{nfw}}(r)= {\varrho_s}{\left( \frac{r_s}{r}\right) \left(
1 + \frac{r}{r_s}\right)^{-2}} \,$}, with the parameters given in the
legend.  Further, the upper panel of Fig.~\ref{fig:sigma} shows the
line--of--sight velocity dispersion profiles for two different
samples, calculated for moving bins containing 20 GCs. The dashed line
shows $\sigma_{\rm{los}}(R)$ for all 87  GCs in the lower panel
(i.e.~all ICGCs and  the GCs near NGC\,1399 but excluding 
those within $1.5\,d_{25}$ of NGC\,1404). For $R\ge10\arcmin$, it
corresponds to the profile shown in Fig.~3 of B07+. The second profile
(thick solid line) was calculated after discarding all ``Class\,B''
velocity measurements (shown as small symbols in the lower panel) and
the most extreme ``Class\,A'' velocity (marked with a square in the
lower panel). This sample of ``high--quality'' NGC\,1399/ICGC
velocities comprises 60 objects. The data points show this sample
divided into three independent bins of 20 GCs.  \par Between 60 and
100\,kpc, the dispersion profile of the ``high--quality'' GCs has a
much gentler slope than the one found when using all GCs, and it
continues to decline, even where the full sample shows an increasing
dispersion. We suggest that the \emph{sharp decrease} of the velocity
dispersion in between 75 and 100\,kpc reported by B07+ is mainly due
to one extremely low (``Class B'') velocity at about 11\arcmin{}
which, when included in the moving bin, leads to large dispersion.\par
The \emph{sudden rise} at $\sim\!150\,\rm{kpc}$, on the other hand, is
caused by about six clusters, five of which have ``Class\,B'' velocity
measurements. Two of these (marked by diamonds in the lower panel of
Fig.~\ref{fig:sigma}) have surprisingly blue colours ($C\!-\!R= 0.84$
and $0.88$), which are unusual for GCs.  The deviant ``Class\,A''
object ( marked by a square in Fig.~\ref{fig:sigma}) {} has a velocity
of $607\pm9\,\rm{km\,s}^{-1}$. Such a large offset (of the order
$800\,\rm{km\,s}^{-1}$, corresponding to four times the local
velocity dispersion) from the cluster mean velocity is difficult to
explain for a bound GC.  Under the assumption that this object is
bound, and that the observed radial velocity equals the pericentric
velocity, we estimate the apogalactic distance (using the halo model
of Fig.\ref{fig:sigma},  dashed line) to be of the order $r_{apo}
> 1\,\rm{Mpc}$. For the more massive halo (dotted line), which appears
to be at variance with the data, one still obtains $r_{apo} \simeq
0.5\,\rm{Mpc}$.  From its photometry ($C\!-\!R=1.13, m_R=21.06$),
we derive, using $m\!-\!M= 31.40$ and $V\!-\!R=0.5$, an absolute
magnitude of $M_V=-9.87$. With the Harris \&
Harris~(\cite{harrisharris02}) colour--metallicity relation, we obtain
$[\rm{Fe}/\rm{H}]=-1.7$. 
\section{Results and Concluding Remarks}
The red (metal--rich) GCs seem to be kinematically associated with
their respective host galaxy or with NGC\,1399.  The population of
ICGCs is predominantly blue, as expected from the shallower number
density profile of the metal--poor NGC\,1399 GCs.  For the blue GCs,
it is harder to determine whether a given GC ``belongs'' to a minor
Fornax member or if it is part of the ICGC/NGC\,1399 population.  We
conclude that the vast majority of the blue ICGCs in fact belongs to
the very extended NGC\,1399 GCS.  Out to at least 165\,kpc, the
velocity dispersion profile of the GCs is consistent with the mass
profile derived from the dynamics of the NGC\,1399 GCS within
80\,kpc.\par We propose to distinguish between intra--cluster GCs
i.e.~GCs which are found at large distances from the galaxies in a
cluster (i.e.~a geometrical classification) and \emph{vagrant} GCs,
which are characterised by radial velocities suggesting that they are
not bound to any galaxy in particular, but rather belong to the galaxy
cluster as a whole.  The example of Fornax illustrates the
difficulties one faces when trying to identify vagrant clusters. The
detection of stripped GCs is made difficult by the richness and extent
of the NGC\,1399 GCS and its high velocity dispersion.  For NGC\,1404,
which has a velocity of about $500\,\rm{km\,s}^{-1}$ w.r.t.~NGC\,1399,
but a (projected) distance of only $\sim\!50\,\rm{kpc}$, it will be
extremely hard to distinguish between a superposition along the
line--of--sight and stripping. NGC\,1387, at about 100\,kpc has a
systemic velocity which is just $\sim150\,\rm{km\,s}^{-1}$ lower than
that of NGC\,1399, making it hard to single out stripped GCs, although
the photometry of Bassino et al.~(\cite{b06a}) suggests their
presence.  We suggest that a dynamical study of the
NGC\,1399/NGC\,1404 and NGC\,1387 region might yield evidence for
tidal structures.
\begin{acknowledgements}
Y.S.~acknowledges support from a German Science Foundation Grant
(DFG--Projekt HI-855/2). T.R.~acknowledges support from the Chilean
Center for Astrophysics, FONDAP No.15010003.  We acknowledge the usage
of the HyperLeda database \texttt{(http://leda.univ-lyon1.fr)}.
\end{acknowledgements}

\end{document}